\documentclass[10pt,twocolumn]{article} 
\usepackage{dcolumn}
\usepackage{multirow}
\usepackage{graphics}
\usepackage{ieee}
\usepackage{times}
\usepackage{url}
\usepackage{xspace}

\newcommand{\gses}{$G_\mathsf{ses}~$}
\newcommand{\gwop}{$G_\mathsf{wop}~$}

\def\urb{$\mu$RB}

\begin{document}

\title{End-User Effects of \\ Microreboots in Three-Tiered Internet Systems}

\author{George Candea and Armando Fox \\  \tt \{candea,fox\}@cs.stanford.edu}

\maketitle

%---------------------------------------------------------------------------
\begin{abstract}

Microreboots restart fine-grained components of software systems
``with a clean slate,'' and only take a fraction of the time needed
for full system reboot.  Microreboots provide an application-generic
recovery technique for Internet services, which can be supported
entirely in middleware and requires no changes to the applications or
any a priori knowledge of application semantics.

This paper investigates the effect of microreboots on end-users of an
eBay-like online auction application; we find that microreboots are
nearly as effective as full reboots, but are significantly less
disruptive in terms of downtime and lost work.  In our experiments,
microreboots reduced the number of failed user requests by 65\% and
the perceived downtime by 78\% compared to a server process restart.
We also show how to replace user-visible transient failures with
transparent call-retry, at the cost of a slight increase in
end-user-visible latency during recovery.  Due to their low cost,
microreboots can be used aggressively, even when their necessity is
less than certain, hence adding to the reduced recovery time a
reduction in the fault detection time, which further improves
availability.

\end{abstract}

\section{Introduction}

Transient faults account for a large fraction of failures in today's
Internet systems and production software in
general~\cite{murphy:reliability, gartner:sustainable}; even
mainframe-class operating systems are not immune to such
transients~\cite{sullivan:defects}.  Running out of memory or file
descriptors, Heisenbug-triggering load spikes, deadlocks, performance
degradation due to unexplained interactions between subsystems,
etc. are just few examples of what Internet service operators face on
a regular basis~\cite{ht:reboot, chou:FT}.

Reboots have been shown to be an effective way to cure many such
transients, even in critical software~\cite{GAO:patriot,
washpost:raptor}.  Full-system reboots, however, can be
expensive~\cite{ebay:power} both in terms of downtime and amount of
disruption; to mitigate this, we introduce the concept of a
microreboot.

In this paper, we demonstrate that microreboots can be used to improve
the availability of applications hosted on a rich middleware platform.
A microreboot is a restart of a subset of fine-grained (smaller than a
process) software components in a running system.  In this paper, the
system in question can be any Java 2 Enterprise Edition (J2EE)
application running on an open-source J2EE application server that we
have augmented with fault injection, instrumentation, and the ability
to microreboot individual beans (application components) as well as
functional subsystems such as the Web server tier and Java Servlet
Pages.

Microrebooting reflects the emphasis on improving availability by
lowering mean time to recover (MTTR); availability is commonly thought
of as $MTTF/(MTTF+MTTR)$.  With respect to interactive services,
lowering MTTR not only improves the user experience of the service and
the users' perception of service availability~\cite{xie:modeling,
fox:mttr}, but also serves as a leverage point for applying aggressive
statistical-anomaly-based failure detection~\cite{ling:ssm,
chen:pinpoint}.  In our case, we also demonstrate that microreboots
can reduce the number of users impacted by a particular transient
failure and the amount of work they lose.

Because our approach is based on observation and control at the
middleware layer, it is application-generic and requires no \emph{a
priori} knowledge of application structure.  This addresses the fact
that today's services are heterogeneous and dynamic, encompassing many
vendors' hardware and software components that evolve rapidly and often
turbulently, resulting in a main challenge in maintaining dependability
of those services~\cite{gray:challenges}.

\subsection{Contributions}

The main contributions of this paper are to demonstrate the efficacy of
microreboots as a technique for improving the availability of
distributed interactive applications and to circumscribe the types of
failures and applications for which microreboots are effective.
Specifically:

\begin{itemize}

\item We identify a user-centric availability metric for
      characterizing the availability of Internet services.  This
      metric reflects the observation that not all types of user
      interactions contribute equally to user-perceived system
      availability.

\item We augment an off-the-shelf Java application server with the
      ability to microreboot individual components in unmodified J2EE
      applications. We show that microrebooting one or a small number
      of components is just as effective as restarting the entire
      application or server (full reboots are currently the most
      common method in use for recovering from
      transients~\cite{ht:reboot,oracle:comm}), but that
      microrebooting is faster, leading to lower recovery time.

\item We show that, since microreboots are less disruptive, using them
      for recovery can reduce the number of end users who actually
      experience the failure and allows some transient failures to be
      masked by additional request latency.  At the same time, the
      total amount of work lost during recovery is reduced.

\item We identify specific cases in which microreboots do \emph{not}
      result in satisfactory recovery, explain why this is the case,
      and propose concrete changes to the middleware (not to the
      applications themselves) to remedy this.

\end{itemize}

Microreboots separate the concern of recovery from that of diagnosis
and bug finding.  When an online system fails, downtime is expensive
and the first priority is to restore service by any means necessary.
Identifying and fixing the root cause of the transient failure is a
separate effort, and we do not claim that microrebooting offers any
insight into doing these things, nor that it is more than a
``temporary fix'' to recovering from the transient.  We expect
production systems to have thorough logging mechanisms that will allow
developers to fix root causes; microreboots improve availability (by
lowering MTTR) without changing reliability (reflected in MTTF).

Therefore, in this paper we try to isolate the effects of
microrebooting as a recovery procedure as much as possible.  In
particular, our microbenchmarks trigger recovery actions directly,
rather than injecting faults and waiting for fault detection to
trigger recovery.  We recognize that reducing fault detection time is
critical, and we explain how microreboots have the potential to enable
the use of promising new approaches to fast and aggressive
detection. Similarly, we recognize that understanding \emph{what} to
microreboot when a failure occurs is important; we have addressed that
problem elsewhere~\cite{candea:afpi} and we use the results of that
technique to drive the experiments in this paper.

In Section~\ref{sec:urbs} we describe the microrebooting approach and
motivate its use for three-tier Internet applications.  We then
describe the changes we made to enable microreboots in JBoss, an
open-source J2EE application server (middleware platform).
Section~\ref{sec:setup} describes our experimental setup, sample
application, and metrics.  Section~\ref{sec:results} presents
experimental results, using trace playback and induced recovery, to
support our claims.  We discuss implications of the approach in
Section~\ref{sec:discussion}, survey related work in
Section~\ref{sec:related}, and then conclude.

\section{What Are Microreboots?}
\label{sec:urbs}

In this section we explore the conditions under which reboot-based
recovery is feasible and describe the concept of a microreboot.  We
present the platform chosen for our work, how it satisfies the
conditions for reboot-based recovery, and conclude with a description
of our implementation of the microreboot mechanism.

\subsection{Reboot-based recovery}
\label{sec:rb-reco}

Chandra and Chen~\cite{chandra:whither} and Lowell and
Chen~\cite{lowell:transparency} formulated an approach to
application-generic recovery (i.e.\ recovery without
application-specific knowledge) based on checkpointing, and demonstrated
that relatively few existing applications could be successfully
recovered by this approach.  They studied both Unix-style monolithic
applications such as {\em vi\/} and large open-source Internet service
components such as MySQL and Apache.

However, part of the appeal of rebooting as a recovery
technique~\cite{candea:RR} is precisely that it \emph{discards}
corrupted transient state that might itself be the cause of the
failure or whose cleanup may be necessary in order for recovery to
succeed.  Therefore we expect that replacing recovery with
rebooting---which is logically equivalent to restarting from a
checkpoint that is the start state of the component---is more likely
to work, \emph{assuming it is safe to try}.  To ensure that it is safe
to try, we must consider three environmental conditions:

\begin{enumerate}

\item Boundary: There must be a clear boundary around what is being
rebooted, i.e., it should be possible to indicate unambiguously what
state will be lost, what resources released, what loci of control
returned to their start state, etc.  For example, in the case of a
process, the boundary is typically the process's heap and any kernel
data structures or resources being maintained on the process's behalf.

\item Loose coupling: if the entity being rebooted is part of a
distributed system, other entities that communicate with it must be
able to tolerate the reboot event as normal.  For example, in a
distributed system, calls to an RPC server that has failed and is in
the process of recovering could be stalled or temporarily rerouted to
a failover RPC server.

\item Preserving state and consistency: To avoid data loss, we must be
sure that all state visible outside the component's boundary is either
soft or discardable state or is committed to a separate persistent state
store (which presumably has its own recovery procedures in case of
failure).  For example, UDP multicast tree information is discardable
soft state whose reconstruction is explicitly part of the corresponding
routing protocol.

\end{enumerate}

Note that these conditions do not guarantee that rebooting will be a
successful recovery method, only that it will not result in a change in
application semantics (e.g., loss of data or loss of consistency).

Analogous to rebooting, a microreboot is a logical restart of an
application component that may be finer-grained than a process, but
the same requirements apply.  Architectures that have these
properties, such as Microsoft's .NET and Sun's Java 2 Enterprise
Edition (J2EE), provide an excellent platform for studying
microreboots.  For the work presented here we used J2EE; in the next
section we briefly review the J2EE programming model and why J2EE
applications meet our requirements.

\subsection{The synergy between \urb~and J2EE}

Most Internet applications are deployed in a ``three tier''
arrangement~\cite{jacobs:bea,cecchet:ejb} that makes state management
explicit.  The presentation tier consists of stateless Web servers
that handle and demultiplex incoming connections.  The application
logic tier runs the code that constitutes the application.  Finally,
the persistence tier stores state that is expected to survive across
requests, whether per-user or across all users.

\begin{figure}[!h]
   \centering
   \includegraphics{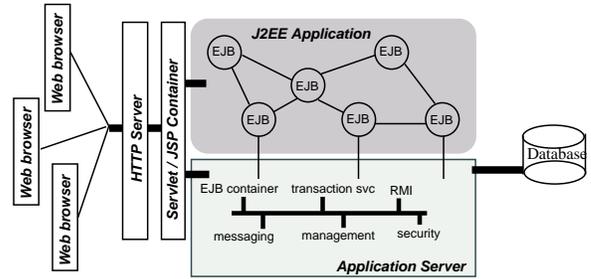}

   \caption{\small Architectural diagram of JBoss.  End-user requests
            enter via the HTTP front end and are serviced by a subset
            of the EJB components.  The database provides persistence
            for application and sometimes session state.}

   \label{fig:arch}
\end{figure}

J2EE~\cite{sun:j2ee} is a model for constructing the application logic
tier of such applications and lends itself well to a microreboot-based
failure management approach.  J2EE applications are constructed from
reusable Java modules, called Enterprise Java Beans (EJBs).  Beans run
in the managed environment of an application server, which provides
\emph{containers} in which the beans are instantiated and run,
provides naming/directory/authentication services for inter-bean
communication at runtime, etc. (see Figure~\ref{fig:arch}).

To run a J2EE application, one must boot the operating system, start
the J2EE application server, start any necessary additional components
required by the application (e.g.,\ a database used for persistent
state storage and the Web server front-ends), and finally ``deploy''
the application on the application server, i.e.  instantiate each EJB
in its container and allow the application to begin accepting requests
from the Web servers.  Once the application is running, Java threads
are mapped to incoming Web requests, and several EJBs may be called
during servicing of a given request.  Thus EJBs are really akin to
event handlers: each EJB does not have a separate locus of control,
rather a single thread ``shepherds'' the user request through multiple
EJBs.  With this description, we can now see how the safe-reboot
requirements from Section \ref{sec:rb-reco} map onto J2EE
applications.

Item (1), clear boundaries around microrebootable components, maps
onto J2EE in at least two different ways.  First, each EJB is a
well-circumscribed entity that can be microrebooted by undeploying it
and redeploying it.  Second, the Web server processes that dispatch
incoming HTTP traffic to EJBs are also self-contained and can be
restarted.  We describe in detail how this is done in
Section~\ref{sec:microrebootsinjboss}.

For item (2), loose coupling, observe that since inter-bean calls are
mediated by the application server, we can modify the application
server to intercept on those calls.  In particular, if an EJB is in
the process of being microrebooted, we can stall calls to that EJB
rather than allowing them to experience an error as a result of the
callee EJB being unavailable.

Item (3), maintenance of persistent state, essentially falls out of
the J2EE application model.  J2EE applications manipulate two kinds of
state that are visible from outside an EJBs boundary.  Persistent
state, such as user profiles and static content, is stored in a
traditional RDBMS and accessed via JDBC connectors.  RDBMSs are well
known for having robust, if not always fast, recovery procedures that
provide strong data integrity guarantees.  The second kind of
non-transient state is session state, which is tied to the maintenance
of a particular user's session (a set of related interactions with the
service).  Since HTTP is stateless and most browsers provide only
cookie management facilities, any nontrivial session state must be
managed by the service.  J2EE provides an abstraction called a
\emph{stateful session bean} that preserves session state across
invocations, but the precise implementation of this abstraction varies
among J2EE application server implementations.  As we describe in
Section~\ref{sec:results}, a deficient implementation of this feature
could cause microreboots to change application semantics; in that same
section we propose a solution to this problem that leverages existing
research and does not require changing individual applications.

In summary, the J2EE application model is a good fit for the
requirements of microrebooting, with the possible exception of the
management of session state, which we will return to in
Section~\ref{sec:results}.  We now describe the particular J2EE
application server implementation that we augmented for the work in
this paper.

%---------------------------------------------------------------------------

\subsection{Microreboots in JBoss}
\label{sec:microrebootsinjboss}

We built upon the popular JBoss J2EE server~\cite{jboss:docs}, because
it is open source and because its performance and features compare
favorably with proprietary closed-source
offerings~\cite{metagroup:j2ee}.  Its use in production environments
is increasing rapidly, having had over 3 million downloads this year.
We instrumented JBoss in a number of ways; a description of the early
changes we made appeared in~\cite{candea:jagr}.  In this paper we
focus on how we enabled the application server for microreboots.

Our basic microrebootable component of J2EE applications is the EJB.
In the same way an OS kernel does for processes, JBoss maintains for
each active EJB a rich set of metadata.  Some of the items include the
name under which this component is known to other parts of the
application, the Java class implementing its functionality, the type
of EJB (session, entity, etc.), whether the bean requires
transactional support, along with references to other beans that this
EJB might call and references to the resources required by the EJB.

JBoss already includes a mechanism for cleanly ``shutting down'' an
EJB; our microreboot mechanism builds upon that.  A Java class
runnable as an EJB must implement the standard EJB interface.  When an
EJB is created, its constructor {\tt ejbCreate()} gets invoked and,
when it is destroyed, {\tt ejbRemove()} allows it to clean up prior to
deletion.  When the application activates an EJB to process a request,
it invokes the {\tt ejbActivate()} method; when the EJB is
disassociated, its {\tt ejbPassivate()} method is called.  When
microrebooting an EJB, our version of JBoss discards all metadata
associated with the bean and resets the corresponding entries in the
server structures that keep track of the bean; if the EJB was involved
in any ongoing transactions, those transactions are aborted.  To
restart the EJB, we simply use the existing deploy mechanism to
reinstantiate it.

In the simple call model, whenever an EJB wants to invoke another
EJB's method, it looks up the target EJB by name in the Java Naming
Directory (JNDI) and uses the Java class resulting from the lookup to
make the invocation (similar to the way RPC stubs work).  Since doing
a lookup on every call is expensive, JBoss provides the caller with a
proxy on the first lookup, which then handles all subsequent calls
without interacting with JNDI.

A problem can arise when a recovering EJB is called by another EJB or
a servlet; all interactions between EJBs, however, are controlled by
the application server.  When a bean is microrebooted, other
components can see the effect of this in one of two ways: either the
bean is not currently registered in JNDI (which would result in a
failure when the bean reference is looked up), or it is not available
for processing requests.  In either case, we arrange for the calling
proxy to receive a {\em RetryLater(t)} exception, where $t$ is the
estimate, in milliseconds, of how much longer it will take for the
callee to recover.

EJBs run inside bean containers, and it is these containers that are
in charge of making the various calls outside the bean.  Our modified
JBoss container catches {\em RetryLater(t)} exceptions, pauses for
approximately $t$ milliseconds, and retries the call in the hope that
the target bean has recovered.  If the call succeeds after a
predetermined number of tries, the original bean code sees a
successful call and has no knowledge that the target bean had actually
failed and recovered in the meantime.  Otherwise, the container throws
an exception to the caller.  These transparent retries allow us to
mask transient failures from callers, as will be shown in
Section~\ref{sec:illusion}.  Such masking makes transient failure an
acceptable mode of operation.

A major issue when transparently retrying calls is idempotency.  In
the present case, however, JBoss guarantees invocations to be atomic:
if the call can be made to the component, it goes through, otherwise
the {\em RetryLater} exception is thrown; this preserves JBoss's
regular call semantics.  As expected, calls that were in-progress at
the time of the microreboot will fail in exactly the same way they
would fail if the EJB crashed, had a bug, etc.

\section{Experimental Setup and Metrics}
\label{sec:setup}

In this section we describe the testbed for our work; we describe
briefly our sample J2EE application, the client emulator used to
generate load on the application, and then conclude with a definition
of the metric used to quantify the benefits of microreboots.

In our hardware setup we tried to mimic what would be typical of a
small Internet service.  We use Linux RedHat 9.0; JBoss and the Web
tier run on an AMD Athlon XP 2600+ PC with 1.5 GB RAM, and the
database (MySQL Max 3.23) on another identical node.  We use the Sun
HotSpot JVM 1.4.1. and allocate it 1 GB of RAM through command-line
arguments.  Our client emulator, described below, runs on on a dual
Pentium III (2$\times$866 MHz) with 1 GB RAM.  All machines are
interconnected by a 100 Mbps Ethernet switch.

\subsection{eBay-like test application}

RUBiS~\cite{cecchet:ejb} is an open-source web-based auction
application, developed at Rice University and modeled on eBay.  It
offers selling, browsing and bidding.  It distinguishes three
kinds of users: visitor, buyer, and seller, with buyer and seller
sessions requiring login. A buyer can bid on items and consult a
summary of their current bids, rating and comments left by other
users. Seller sessions require a fee before a user is allowed to put up
an item for sale.  The seller can specify a reserve (minimum) price for
an item.  RUBiS contains 582 Java files and about 26K lines of code; it
uses MySQL for the database back end and stores 7 tables.  In the
default configuration, RUBiS has about 33,000 items for sale,
distributed among eBay's 40 categories and 62 regions. There is an
average of 10 bids per item, or 330,000 entries in the bids table. The
users table has 1 million entries.

\begin{figure}[!ht]
   \centering
   \includegraphics{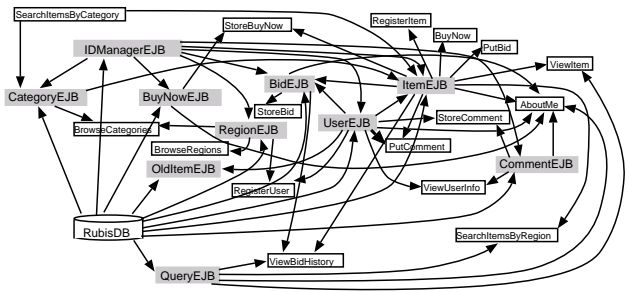}
   \caption{\small {\bf Fault propagation map} for RUBiS, obtained
            with AFPI.  Shaded boxes represent EJBs, clear boxes
            represent servlets (which are inherently stateless).  An
            edge from $A$ to $B$ indicates that a fault in $A$ was
            observed to propagate to $B$; microrebooting $A$ entails
            microrebooting the transitive closure of $A$ over this
            fault propagation map.}
   \label{fig:rubis_fmap}
\end{figure}

We obtained a description of the failure dependencies between RUBiS's
components using automated fault-propagation inference
(AFPI)~\cite{candea:afpi}.  As can be seen in
Figure~\ref{fig:rubis_fmap}, the majority of such dependencies in
RUBiS are between the stateless servlets and EJBs; the only beans that
can propagate faults to other beans are {\tt IDManagerEJB}, {\tt
ItemEJB}, {\tt CategoryEJB}, and {\tt UserEJB}.  AFPI information is
collected during a completely automated fault-injection campaign that
requires no \emph{a priori} knowledge of the applications' structure
or semantics.  In this paper, however, we focus on \urb-ing as a
technique, not on how we might use \urb{}s in a production system or
what policies we might devise based on AFPI-generated graphs.

For RUBiS, a simple recovery policy could be based on the fact that
there is a known mapping from the URL being accessed to the action
being taken (and hence the EJBs being touched).  For instance, if a
failure is seen on {\small
http://ejb\_rubis\_web/servlet/BrowseCategories}, then we know that
something in the path starting at the BrowseCategories servlet has
gone wrong, and the system should therefore automatically \urb~the
corresponding components.

\subsection{Client emulator}

Our client emulator is a modified version of the load generator that
ships with RUBiS.

We describe the workload of the simulated clients using a state
transition table $T$ that has the client's possible states as rows and
columns; these states correspond naturally to the various operations
possible in RUBiS, such as {\em Register}, {\em
SearchItemsInCategory}, {\em PutBid}, etc. (27 in total).  In addition
to the application-specific states, we also have two states
corresponding to the user hitting the back button (Back) and
spontaneously deciding to end his/her session (End).  $T(S_a, S_b)$
represents the probability of a client transitioning from $S_a$ to
$S_b$; e.g., $T(\mathrm{ViewItem}, \mathrm{BuyNow})$ describes the
probability we associate with the user clicking on the ``Buy Now''
button while viewing an item's description.

The client emulator uses this table to automatically navigate the web
site; when in a given state $s$, it will randomly choose the next
state based on $T(s)$ with the requested probability; it then
constructs the URL for this state and ``clicks'' on it.  The table $T$
also has a column for how long a user waits inbetween clicking from a
certain state to the next; in our experiments however we set this wait
time to zero.  Unlike a real user, our emulator will therefore
initiate the next HTTP request as soon as the current request
completes (whether successfully or not). The emulator uses one thread
per simulated client.

We classify responses from the server as correct or incorrect.
Correct responses are used to compute the server's goodput (throughput
of correct responses per second); we will describe our other metrics
in more detail later.  A response will be classified as incorrect if
it results in a network-level error (cannot connect to server, etc.),
an HTTP 4xx or 5xx error code, or an HTML page containing particular
keywords that we know to be indicative of application errors.  Clearly
this is not fully application generic; we have successfully detected
all error pages in RuBiS and other J2EE applications with no false
positives by searching for ``error'', ``failed'', and ``exception'' in
the reply HTML, but this assumes none of the users is selling an item
that would match these searches.

%---------------------------------------------------------------------------

\subsection{Metrics}

\begin{figure*}[!ht]
   \centering
   \includegraphics{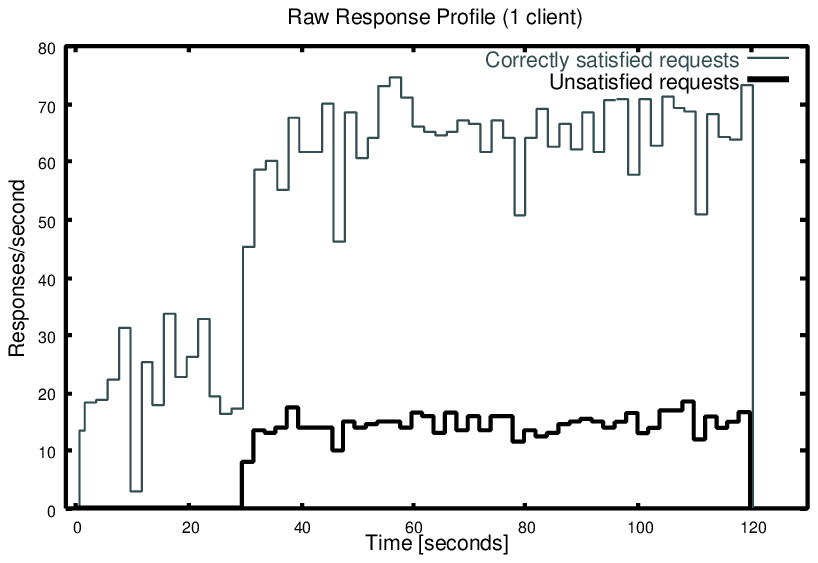}
   \includegraphics{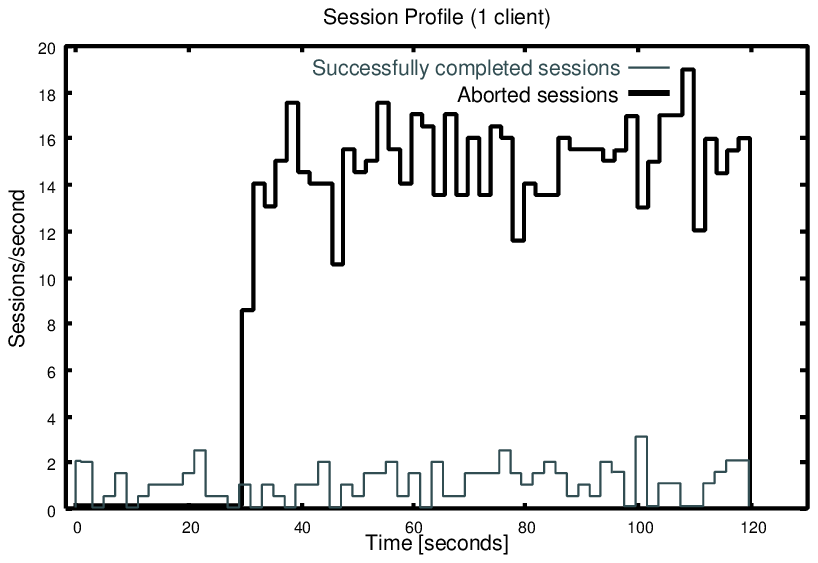}
   \caption{\small {\bf Goodput anomaly}: We induced a failure
            (unrecovered) in QueryEJB at $t=30$, thus causing certain
            queries against the database to fail.  On the left we show
            a request profile (both good and failed requests); on the
            left we show a profile of the sessions (both the ones that
            completed and those that failed).  Witness the goodput
            anomaly: in the face of failure, the raw goodput increases
            simultaneously with the rate of unsatisfied requests.  The
            session profile, however, shows that the number of aborted
            sessions goes from zero to an average of 14 aborted
            sessions/second.  Users start a new session after the
            previous one has failed, so the increased number of
            aborted sessions brings about an increase in the session
            initiations---in real services, this could constitute an
            unwelcomed load spike for certain types of services, such
            as user login and authentication.}

   \label{fig:1cli_fault}
\end{figure*}

A typical interaction of a client with the web site proceeds as
follows: client goes to the homepage, browses around for a while
performing different site actions (searching, etc.), and then decides
to do something that touches the persistent-state database (e.g.,
place a bid, leave a comment for a user, update his/her profile,
etc.).  The DB-touching operation(s) usually require the user to have
logged in.  We assume that all interactions preceding the
persistent-state update are just precursors to the real action (the DB
update is sort of a ``crowning moment''); thus, these interactions
serve no purpose (with respect to the proposed metric) in the absence
of a successful ``crowning moment.''

For the purpose of this paper, we define a session to be a sequence of
interactions with the web site that starts at the homepage and ends
with the emulated user either abandoning the site or returning to the
homepage (thus starting a new session).  Note that if something goes
wrong during a session, users often try to logout and log back in
(which typically requires going to the homepage), so it is reasonable
to define a session as the sequence of URLs bracketed by accesses to
the homepage.  This definition relies on user behavior to infer when
s/he is done with the site, rather than trying to understand
application semantics.

A simple approach to measuring the effect of downtime on end users
would be to measure goodput (number of requests completed
successfully) under partial-failure conditions, averaged across all
clients.  This is usually how
performability~\cite{meyer:performability} is measured---the amount of
work successfully completed over a period of time in the presence of
partial failures.

\begin{figure*}[!ht]
   \centering
   \includegraphics{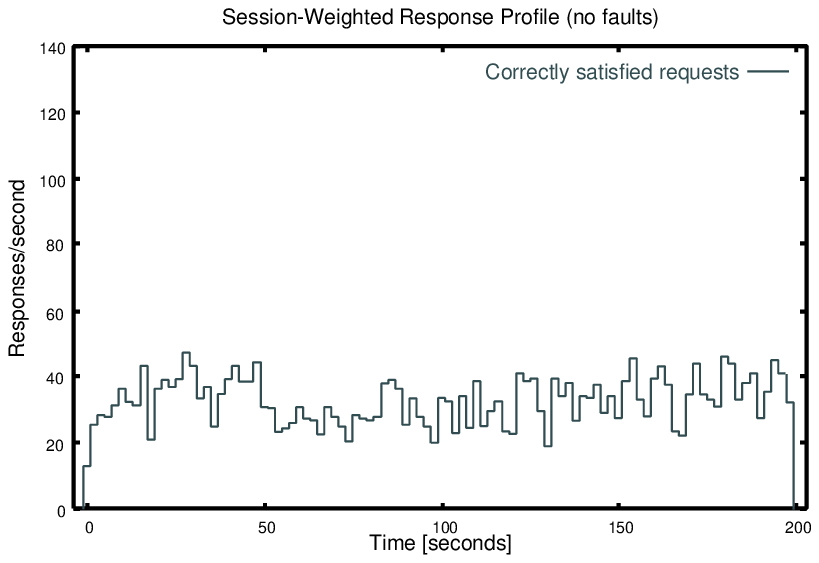}
   \includegraphics{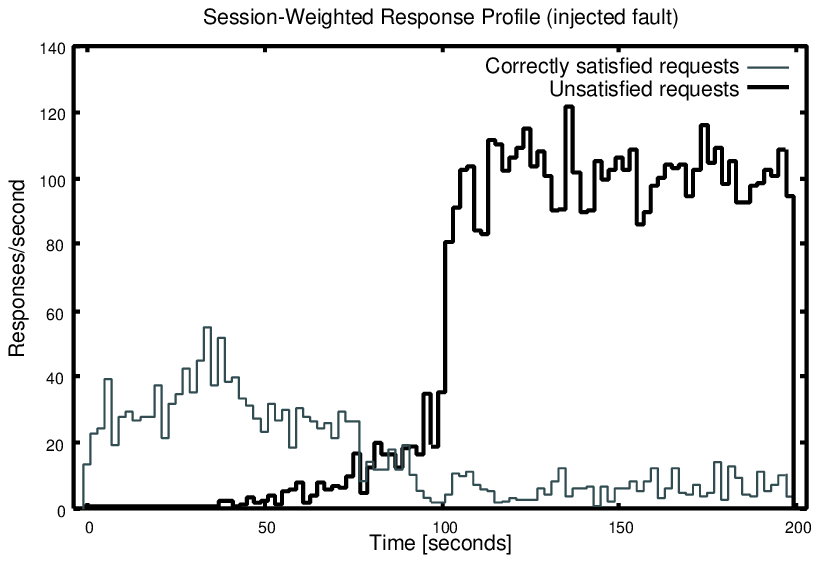}

   \caption{\small {\bf Session-weighted goodput}: On the left we see
            a run with no failures; on the right,
            we inject a permanent/unrecovered fault in QueryEJB at
            $t=100$ sec.  Notice how this causes unsatisfied requests
            to show up even ``before'' $t=100$ sec, because operations
            done as part of sessions that started at $t<100$ and {\bf
            failed} at $t\geq100$ are marked as failed by the \gwop
            metric.  Even though these individual requests were
            satisfied, the usefulness of satisfying them is lost
            because of the session failure.  We also notice that the
            throughput of satisfied requests starts dropping prior to
            $t=100$, for similar reasons: some sessions starting in
            that interval end up failing. }

   \label{fig:Gwop_20_clients}
\end{figure*}

\begin{figure*}[!ht]
   \centering
   \includegraphics{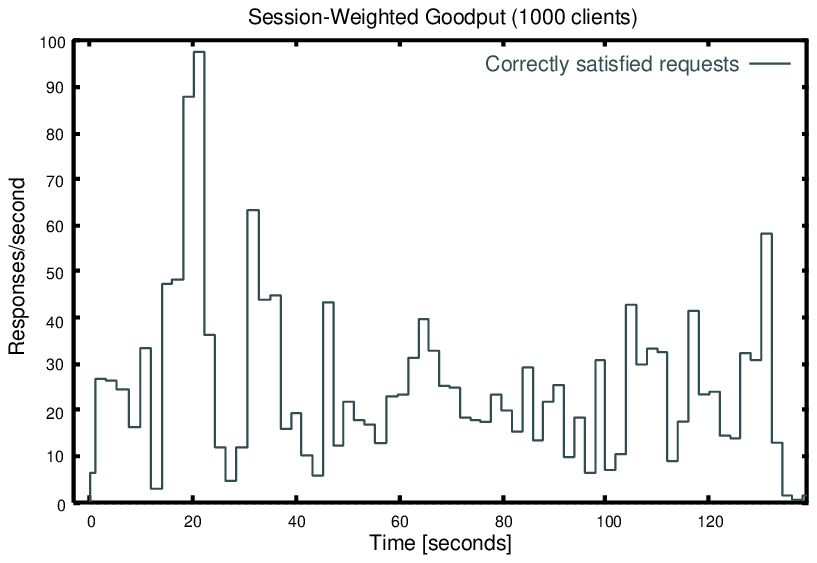}
   \includegraphics{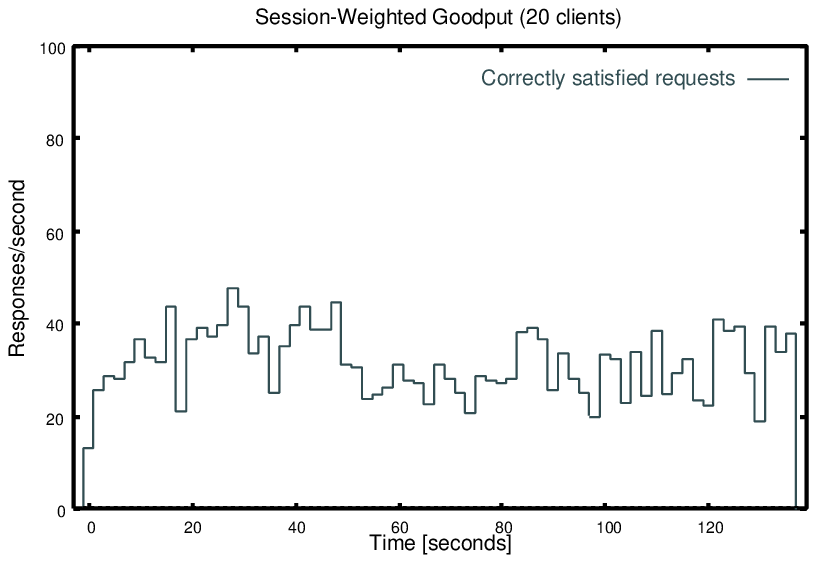}
   \caption{\small {\bf Sweetspot:} On the left we show the
                   session-weighted goodput for a 1000-client load; on
                   the right we show the same for a 20-client load.
                   We empirically determined that the experimental
                   results are most consistent for 20 concurrent fast
                   clients.  For higher numbers than that, our
                   single-node application server starts thrashing and
                   throughput becomes erratic and slightly lower on
                   average, as seen in the left graph.  Moreover,
                   inspecting traces of 20 clients for correctness is
                   considerably easier than 100 or 1000 clients.}
   \label{fig:workload}
\end{figure*}

But the simple goodput metric fails to distinguish between potentially
long-running DB-touching operations and simple, fast browse-only
operations.  The surprising result of inducing failures during
long-running DB operations is that the goodput actually goes {\bf up}
in the presence of failure, because the user no longer waits for a
long-running operation---it fails right away and the client emulator
moves on to the next (non-DB-touching) operations.  In other words, as
figure~\ref{fig:1cli_fault} shows, the simple goodput metric would
fail to capture that some operations are more ``valuable'' than
others, and executing many ``simple'' operations does not necessarily
compensate for failing to execute a few long-running ones.

Instead we propose two metrics.  The first, \gses, counts the number
of sessions in which all operations completed successfully (i.e. every
operation within the session was failure-free).  In Figure
\ref{fig:1cli_fault} we illustrate this metric for 1 client.  Note
that this is a pessimistic metric: the user may believe s/he has
accomplished useful work during the session, but unless \emph{every}
operation succeeds, the session is not counted as successful.  The
second metric, {\em session-weighted goodput} (\gwop), weighs each
session by the number of operations in it.  Viewed differently, \gwop
measures standard throughput of successful and failed requests
respectively, but whenever a session fails, \emph{all} the operations
in that session are counted as failed.  Unlike \gses, \gwop is able to
capture the fact that when a long session succeeds, the user got a lot
more done than when a short session succeeds.
Figure~\ref{fig:Gwop_20_clients} illustrates this metric.  We will use
primarily \gwop to quantify the effect of microreboots in our
experiments.

\section{Results}
\label{sec:results}

\begin{figure*}[!ht]
   \centering
   \includegraphics{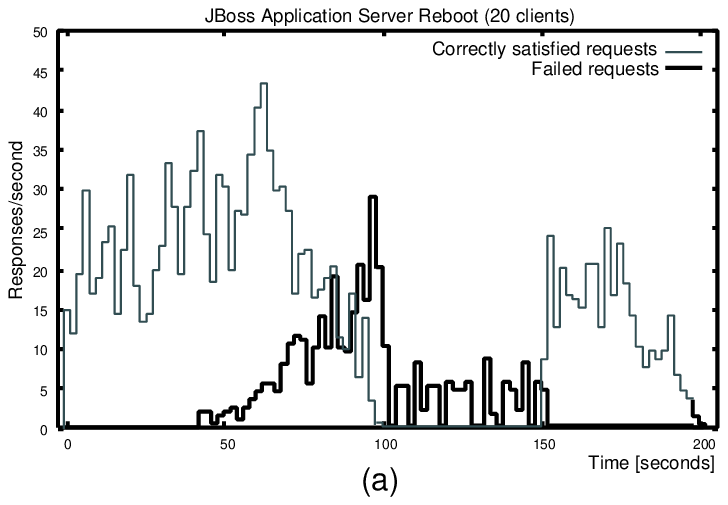}
   \includegraphics{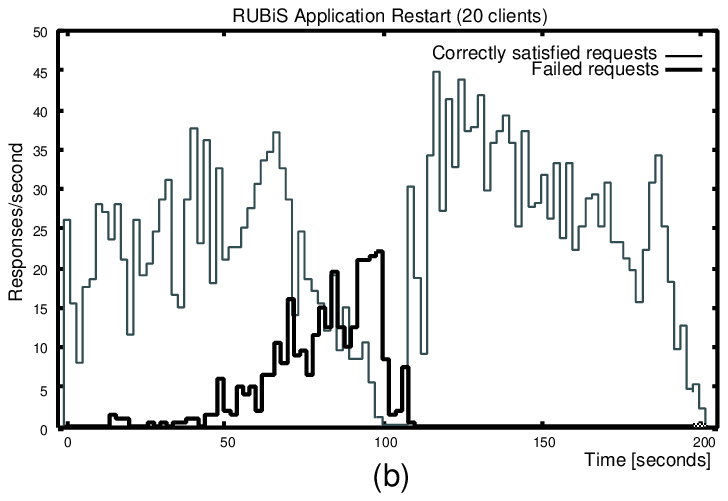}
   \includegraphics{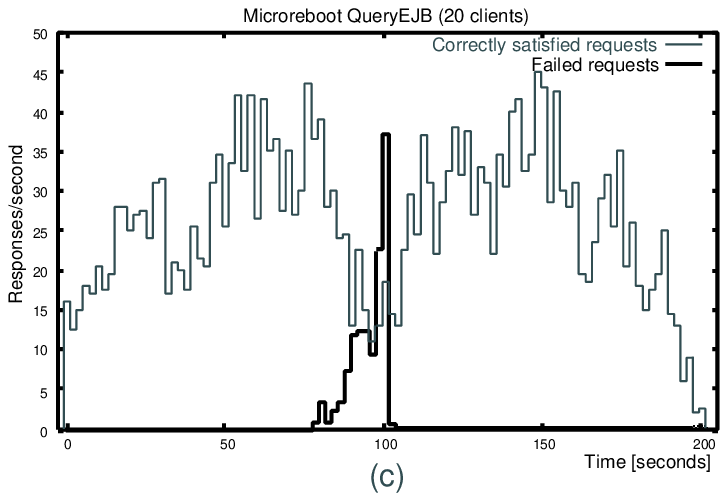}
   \includegraphics{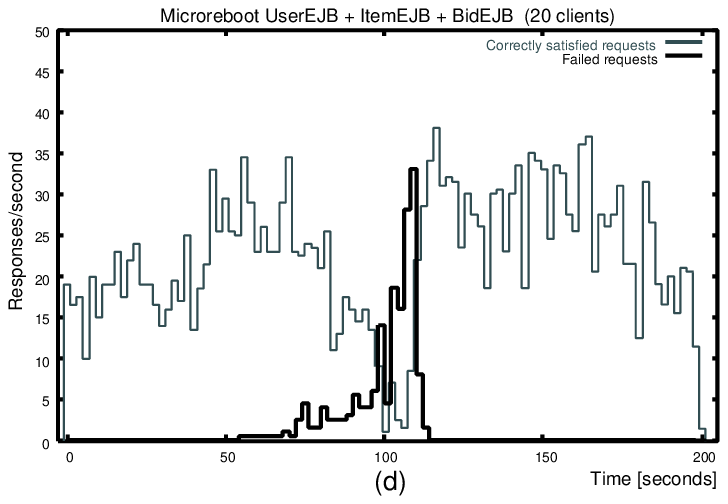}
   \caption{\small {\bf Microreboots vs. other forms of reboot:} Graph
            {\em (a)} depicts the impact on end users of a full JBoss
            server process restart: 713 failed requests over a time
            span of 108 seconds.  Graph {\em (b)} shows a full RUBiS
            application restart: 615 failed requests over 94 seconds.
            Graph {\em (c)} shows the impact of microrebooting one EJB
            with no call retries: 251 failed requests over 24 seconds.
            Graph {\em (d)} shows the impact of simultaneously
            microrebooting 3 mutually-dependent EJBs with no call
            retries: 345 failed requests over 29 seconds.}

   \label{fig:basic_20_clients}
\end{figure*}

In order to experimentally isolate the \urb{} recovery mechanism from
fault detection, we initiated various forms of rebooting in the
application without actually injecting faults.  The aspect of
detecting such faults was the focus of~\cite{candea:jagr}.  We are,
therefore, assuming in all our experiments that the application server
has instantaneously detected the fault and initiated reboot-based
recovery.  Note that the problem of failure detection is orthogonal to
the recovery method used, though in section~\ref{sec:discussion} we
argue that \urb{}s make it potentially much easier to apply certain
kinds of fast failure detection algorithms.

Reboot-based recovery is typical for many failures noticed in deployed
Internet systems, where resource leaks, deadlocks, etc. occur on a
regular basis~\cite{ht:reboot}.  In fact, the original version of
RUBiS/JBoss had a bug that caused it to deadlock when the number of
concurrent users exceeded 10, and we have shown in~\cite{candea:jagr}
that a modified JBoss could automatically recover from this deadlock.
The version of RUBiS used in our experiments incorporates a number of
fixes to enable it to run with many concurrent users.

For most of the results reported here we used 20 concurrent clients
with no think time inbetween successive requests.  Given that a human
user typically spends in excess of 2 seconds between clicks, we
believe the load placed by one of our simulated clients is equivalent
to that of 100 or more real clients.  We did not want to introduce
artificial think time (the way is done, for instance, in the TPC-W
benchmark) because having think time would add one more variable to
the experiment and would not offer any useful insight for \urb{}
experiments.  The reason we settled for 20 concurrent rapid clients is
that it appeared to be the threshold beyond which thrashing and other
side effects would reduce throughput; see Figure~\ref{fig:workload}
for a comparison of throughput for 20 vs. 1000 clients.

An important characteristic of our chosen workload is that it covers
all possible RUBiS operations; experimentally we have determined that,
in runs lasting 1 minute or longer with 20 clients, we routinely
exercised all components.  While this might be surprising for a set of
20 human users, our no-wait-time clients navigate through the site
very rapidly.  The workload we used for the experiments reported here
had an approximate mix of 85\% read operations and 15\% DB write
operations.

In the remainder of this section we will show that \urb-ing is faster
and less disruptive than other forms of reboot, discuss correctness in
the presence of \urb-ing, and conclude with a technique that, in
conjunction with microreboots, can mask transient failures from end
users.

%---------------------------------------------------------------------------

\subsection{Microreboots are fast}

Our first goal was to determine whether the microreboot mechanism we
built into JBoss can indeed reduce recovery time.  We performed four
sets of experiments, comparing full reboot of the JBoss application
server, full reboot of the RUBiS application, microreboot of one EJB
(QueryEJB), and microreboot of multiple dependent EJBs (UserEJB,
ItemEJB, and BidEJB), respectively.  This would be a normal response
to a variety of reboot-curable failures, such as a bean running out of
memory (quite frequent in Java systems) or being hung in a deadlocked
transaction.  We can think of these experiments as attempting to
recover a failure in one of the application components;
Figure~\ref{fig:basic_20_clients} shows the results.

In the four graphs we show profiles both of successfully served
requests and failed requests.  We are particularly interested in
reducing the total number of failed requests (area under the bold
curves in the graphs), as this reflects exposure of the failure to end
users.  We also want to reduce the amount of time during which the
service appears down to any user.  In the case of single-component
microreboot, we reduced the number of failed requests by a factor of
2.84 over server process reboot and by a factor of 2.45 over an
application restart.  The duration of time during which the site was
perceived down by some users was reduced by a factor of 4.5 over
server process restart and by a factor of 3.92 over application
restart.

In graph (d) we show the effect of microrebooting a group of EJBs.
This can be required either because multiple components have failed,
or because there are dependencies between components.  In this case,
we may be reacting to a failure in {\em UserEJB}; as can be seen in
Figure~\ref{fig:rubis_fmap}, the transitive closure of {\em UserEJB}
over the fault propagation graph is the group of beans {\em UserEJB},
{\em ItemEJB}, and {\em BidEJB}.  Based on this information, we
microreboot the three beans together.  By microrebooting the EJBs
instead of restarting the application, we reduce the number of failed
requests by a factor of 1.78 and the downtime by a factor of 3.24.
Table~\ref{tab:improvement} summarizes the results of these
experiments.

\begin{table}[!h]
\small
\begin{tabular}{| c | c | c | c | c | c |}
\hline
Recovery       &  Failed   & Downtime  & \multicolumn{2}{c|}{Improve over JBoss restart} \\
Technique      &  Reqs     & [sec] & Requests & Downtime \\ \hline
%-------------------------------------------------
JBoss restart  &  713      & 108       &  --   &  --  \\
RUBiS restart  &  615      & 94        &  14\% & 13\% \\
1-EJB \urb     &  251      & 24        &  65\% & 78\% \\
3-EJB \urb     &  345      & 29        &  52\% & 73\% \\ \hline
%--------------------------------------------------

\end{tabular}
\caption{\small {\bf Improvement relative to process restart}:
                 Comparison of application restart, one-EJB
                 microreboot, and three-EJB microreboot in terms of
                 failed requests and perceived downtime.}
\label{tab:improvement}
\end{table}

%---------------------------------------------------------------------------

\subsection{Microreboots are less disruptive}

If a recovery method is effective, then users of the recovered system
should be able to resume their work immediately following the
recovery of the system under load.  As
evidenced by the lack of failed requests after recovery
completes (Figure~\ref{fig:basic_20_clients}), the system sustains all
four methods of recovery equally well, and users can continue their
work after microreboots just as in the case of regular reboots.  In
fact, the implementation of whole-application reboot is in effect a
collection of microreboots, because ``rebooting'' the application
consists of undeploying and then redeploying the individual
application components.

As can be seen in Figure~\ref{fig:basic_20_clients}, microreboots have
the potential to be considerably less disruptive than either server or
application restarts.  For example, in graphs (a) and (b) goodput
drops down all the way to zero, while the rate of failed requests does
not increase dramatically.  In both cases, the application loses all
its network connections to the clients; when clients try to establish
new connections, they are refused flat out, since no process is
listening on the corresponding port.  The result is that all users of
the service experience downtime, hence the zero goodput.

When using a \urb~to recover a component, however, we are
effectively partitioning the user population into two groups: those
whose requests require the services of that component and those whose
requests do not.  As seen in graphs
(c) and (d), goodput does not drop to zero even instantaneously,
because non-affected (non-recovering) components can continue to
deliver service while faulty ones are microrebooted.  Users whose
sessions
do not require the recovering component can continue working as if
nothing had failed.  Thus, in addition to recovering faster, microreboots also
offer the opportunity to reduce the impact of failures on  users who
are active at the time of failure.

\subsection{Microreboots and correctness}
\label{sec:sessionstate}

``Correct behavior'' in the face of microreboots is difficult to define,
because by assumption microreboots are useful only because there are
transient bugs in the application or server.  With or without \urb{}s,
such bugs clearly might cause incorrect behavior.  However, we can argue
that the effect of a \urb is no different than the effect of a full
reboot.

First, observe that the result of a particular user request depends
only on the EJB's it calls and on any persistent state that would
affect the way the bean handles the request (whether or not that state
is directly visible outside the bean boundary).  Instances of
stateless beans are by definition indistinguishable from each other,
and in fact application servers simply select an available instance
from a pool of such beans when the bean is needed for processing an
incoming request.  Bean-independent state stored in a transactional
database (e.g., a list of all bids placed by a given user) is not
affected any differently by full reboots vs. microreboots of the
application server: as long as the database provides transactional
semantics, either event is ``serialized'' between two transactions.
The remaining possibility is that the bean is a stateful session bean,
which expects to preserve its state across invocations.  In this case
we need to know where the state is kept, and whether it would survive
a \urb of the bean.

As we stated earlier, management of session state varies across
implementations of J2EE application servers.  JBoss offers two options:
(a) individual EJB's can manage their own session state by explicitly
updating the transactional database; (b) JBoss can transparently manage
session state, which it does by keeping it in RAM with no replication
or backup.  RUBiS happens to use (a), which means all beans' session state
would survive microreboots of the beans themselves.  However, an
application that used (b) would \emph{not} have its session state
preserved across a \urb.  In fact,
we  encountered this problem in PetStore, a J2EE application that models a
simple e-commerce site; after a \urb, all subsequent requests from the
affected (simulated) user  systematically failed, until
the (simulated) user abandoned the session and logged back in (thereby
recreating a new 
valid session-state object), after which
subsequent requests succeeded.  In other words, \urb{}s are \emph{not}
transparent to applications relying on
JBoss's implementation of session state.  (Note, however, that such
applications won't survive full reboots either: in that case, \emph{all}
currently connected 
users would lose their sessions, not just the user whose session beans were
\urb'd.)

We propose that the correct solution to this problem is to manage
session state externally to the EJB's using a mechanism that is much
lighter-weight than a database (for performance and scalability) but
provides strong guarantees of bounded persistence.  In fact,
\cite{ling:ssm} reports on 
a lightweight, RAM-only, replication-based session state storage
mechanism whose contents survive microreboots and whose use does not
compromise overall application throughput.  Note that integrating this
system into our prototype would not require changes to the applications
themselves, only to the application server's implementation of
server-managed session state.  We plan to explore the integration of
this subsystem into our prototype in future work.

%---------------------------------------------------------------------------

\subsection{Low-level retries mask transient failure}
\label{sec:illusion}

When a caller tries to reach a recovering component, that call will
typically fail.  However, we can build in mechanisms for retrying such
calls after the target component has recovered; such retries can be
done in a manner completely transparent to the J2EE application, which
means J2EE programmers do not need to incorporate retry logic in their
applications.  Microreboots coupled with transparent low-level retry
would allow us to transform many transient failures into additional
latency instead of externally visible failure.  The effect of a
failure on users would then simply be a performance hiccup, rather
than the error they would have seen without retry mechanisms.  As
discussed in Section~\ref{sec:microrebootsinjboss}, on calls to
recovering EJBs we provide the same semantics as vanilla JBoss, with
the addition of a {\em RetryLater(t)} exception.

Such retries can be automated and performed transparently at multiple
levels.  The highest level is provided by HTTP 1.1, which has a {\tt
Retry-After} response header, allowing the web server (or our
application server) to instruct the client's browser to retry after a
certain number of seconds.  At the lowest level, calls between EJBs
can be retried if a particular EJB is microrebooting.

Previous studies~\cite{bhatti:response,miller:thinktimes} have found
that, when a user is waiting for an interactive service to respond, a
delay of 8--10 seconds is the threshold after which the user comes to
believe that the request has failed and clicks the Reload or Stop button
(or worse, clicks over to another site).  This suggests that if a site
can recover from a transient failure and retry the failed in-flight
request(s) within 8 seconds, affected users will have the illusion of
continuous uptime---they will see a short delay rather than a failure.
Microrebooting an EJB takes less than 1 second, which permits us to use
call retries to mask EJB failures from most end users.

\begin{figure}[!h]
   \centering
   \includegraphics{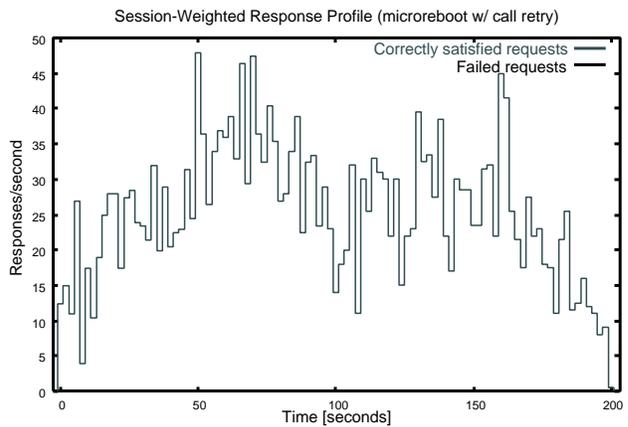}
   \caption{\small {\bf Microreboots with call retry:} QueryEJB is
            \urb-ed at $t=100$; in-flight requests are retried with a
            timeout of 100 msec. Goodput dips around $t=100$ but none
            of the requests fail.  In the case of multi-EJB
            microreboot, the dip would be deeper and wider, but still
            no requests would be perceived as failed by end users
            (unless recovery time exceeds 8 seconds). }
   \label{fig:illusion}
\end{figure}

We built into JBoss the ability to retry inter-EJB calls if the callee
is microrebooting; as described in
Section~\ref{sec:microrebootsinjboss}, a call to a recovering EJB
results in a {\em RetryLater(t)} exception.  Using this facility with
a hardcoded value of $t=100$ msec, we ran an experiment in which we
microrebooted QueryEJB and observed the effect on end users.  We do
not compare these results to application-level reboot, because restart
time exceed the above-mentioned threshold (restarting all of RUBiS
takes on the order of 10-11 seconds with no load, and 20 or more
seconds under normal load).

In Figure~\ref{fig:illusion} we show the session-weighted response
profile for microrebooting QueryEJB with call-level retry.  Notice
that the goodput dips slightly around the time when the microreboot is
made, because the system is now busy retrying in-flight request and
can accept fewer new requests than under normal circumstances.
However, the goodput only drops slightly and, most importantly, no
request fails and hence all sessions can continue unaffected. Since
incoming requests sit in the TCP connection queues, it is conceivable
that if load was much higher and microreboot time longer, the TCP
queues would fill up and connections would be refused (and the site
perceived as down by end users).

%---------------------------------------------------------------------------

\section{Discussion}
\label{sec:discussion}

\subsection{Cheap recovery allows occasional mistakes}

Microreboots enable an approach to self-management in which recovery
is so fast and inexpensive that false positives during failure
detection become less important.  The fact that microreboots are fast
and safe potentially allows us to apply much more sophisticated
failure detection policies, which is important because total recovery
time is often dominated by fault-detection time~\cite{chen:nsdi}.

One promising direction involves using statistical anomaly detection
to infer failures.  Such techniques have been shown to reduce
time-to-detection at the cost of some false
positives~\cite{chen:pinpoint}, and a simplified version of this
approach appears to have been successfully demonstrated in a
state-management layer designed specifically to make reboots fast and
safe~\cite{ling:ssm}.

In general we believe this is an important design trend for robust
systems: fast, cheap recovery mechanisms will blur the line between
``normal operation'' and recovery.  When recovery becomes an order of
magnitude cheaper, it allows one to think differently about how and
when to apply it.  Since microreboots result in only a minor cost in
goodput if applied by mistake, they provide the level of recovery
performance needed to pursue a rationally-aggressive approach of
initiating recovery at the slightest hint of failure.

%---------------------------------------------------------------------------

\subsection{\urb-ing in Internet services}

We believe the \urb~technique is best suited for large scale Internet
services.  The workloads faced by such services consist of
short-lived, mostly-independent requests coming from a large
population of distinct users.  The work that Internet services must do
is generally partitioned into disjoint sets of discrete operations,
and RUBiS reflects this.  The consequence is that, even if a few
requests fail, it is possible for most users to be unaffected.
\urb{}s take advantage of the application's structure to realize this
potential.

Additionally, the underlying protocol (HTTP) and most of the
application logic is stateless and, except for marked, non-idempotent
requests, end-users can safely retry failed requests until they
succeed.  This lets us reboot components in the system, knowing that
any users affected will face only a minor inconvenience.  In fact,
this property makes it useful to recover even from purely
deterministic bugs such as a pathologically malformed request: if
recovery is fast enough, other users issuing non-pathological requests
may still be able to use the 
service.

Many Internet services today use huge in-memory caches in order to
avoid the central database bottleneck (e.g., the servers at a large
Internet portal use 64 GB of RAM just for caching database
queries~\cite{pal:comm}).  Unfortunately, a machine reboot flushes
this cache, and re-warming it can take a long time (transferring 64 GB
from a 40 MB/sec wide-SCSI disk takes at least half an hour), which is
why whole-system reboots are generally avoided.

%---------------------------------------------------------------------------

\subsection{When reboot-based recovery fails}

Rebooting is a correctness-preserving form of restart only to the
extent that no ``critical'' state is lost and no inconsistency
created.  We identified three requirements for allowing safe reboots:
a well-defined reboot boundary, loose coupling between the rebooting
component and its peers, and preservation or state (or preservation of
consistency of state) visible outside the reboot boundary.  We chose
to target three-tiered Internet applications based on thick middleware
because the programming model largely enforces these properties
already.  Monolithic applications, in contrast, typically lack these
properties, and we would not expect microreboots to work without
extensive changes to the applications themselves.

Also, since microrebooting still introduces nonzero recovery latency, it
may be inappropriate for systems with tight real-time constraints; hot
standby with fast failover may be the only acceptable option in those
cases, since even failover to a cold spare may take too long.  Of
course, substantially all Internet services exploit some form of standby
at multiple levels to mask transient failures~\cite{marcus:ha}, but
standby capacity is expensive so standbys are rarely kept idle during
normal operation.  As a result, when failover does occur, the standby is
serving more than its steady-state share of workload; presumably, the
faster the primary is returned to its online condition, the higher total
throughput can be realized.  In fact, the CNN.com meltdown on
9/11/01~\cite{lefebvre:cnn} demonstrated that slow node-level recovery
time can lead to service collapse even when fast failover is in place.
Ideally, if more transient failures could be masked as slight delays to
the user, fast failover would not have to be used as often.
Microreboots are thus orthogonal to failover as a technique, but they
may result in failover being required much less often.

\subsection{Application-generic recovery}

In applying microreboots to the application server rather than by
modifying individual applications, we have attempted to address the
question: What level of effective fast recovery can we do exploiting
just the middleware, and without application-specific efforts?  We
found that microreboots can recover just as effectively from failures
that today are usually cured by whole-system restart or
whole-application restart, but they do so more rapidly.  Moreover, the
shorter downtime during recovery can sometimes be masked by a delay
well within an established ``distraction threshold'' for most users,
allowing the failure to be hidden from them; this is not practical for
the longer recovery times required for whole-application restart.

In our experience we have found few failures from which whole-system
restart recovers but microreboots do not.  Nevertheless, such failures
exist; for example, we observed that under high load, the JVM running
the application server would sometimes run out of file descriptors, or
encounter an internal error, requiring a process restart of the JVM.
We have also encountered a resource leak involving serialized objects
sent over a socket: the object does not get garbage collected even
when our references to it are gone, and  eventually the leaks require a
JVM restart.  Finally, on our version of Linux, we also encountered on
occasion a kernel bug in the swapping code which would trigger under
high memory utilization conditions; in such cases, any memory
allocation (specifically, any call to the {\tt brk} system call)
hangs, and a full system restart is necessary.

Although none of these problems would have been cured by microreboots, 
we still expect microreboots to recover from a significant subset of
those faults that a full reboot would cure.  The strategy we proposed
in~\cite{candea:RR} suggests that restarts be attempted at the finest
granularity, and then progressively encompass more components if the
failure is not cured.  In this sense, microreboots as presented here
are an optimization over full reboots, albeit one that results in
qualitatively less disruptive recovery than full reboots.

\section{Related Work}
\label{sec:related}

Chandra and Chen~\cite{chandra:whither} classified software faults into
three categories.  Environment-independent (EI) or ``Bohr bugs'' are
deterministic and do not depend on the operating environment;
environment-dependent-transient (EDT) or ``Heisenbugs'' are due to
timing or  other 
transient conditions and may disappear if the operation is retried;
environment-dependent-nontransient (EDN) bugs are related to the operating
environment in such a way that immediate retry is {\em not\/} likely to
work, because the environmental condition(s) responsible for the bug
will not have changed enough (for example, a failure due to a memory
leak will persist until more heap space is made available).
There is disagreement regarding whether most bugs remaining in
production-quality software are EDT~\cite{oracle:comm} or
EI/EDN~\cite{chandra:whither}, but reboot-based recovery techniques
address all three categories.

Other projects attempting checkpoint-based recovery have
fared similarly to Lowell and Chen~\cite{lowell:transparency}.
ARMORs~\cite{chameleon} provide a micro-checkpointing 
facility for application recovery, but applications must be (re)written
to use it;  limited protection is provided for legacy applications
without their own checkpointing code.  ARMOR's own fault detection and
recovery middleware does use the microcheckpointing facility, but in the cases
where middleware recovery failed~\cite{whisnant:ree_sift_armor}, it was
because of a corrupted checkpoint caused by an injected fault.
Libft~\cite{huang:libft} provides a C library for checkpointing and
the internal state of an application process periodically on
a backup node, but like ARMOR, it requires applications to be written
specifically to use this feature.

Process pairs~\cite{nonstop} were an early mechanism that combined
resource redundancy and state mirroring to allow failover to a hot
standby, but because they were difficult for programmers to use, they
have had limited impact outside of specialized high-end systems.
Transactions~\cite{GRAY_81} have enjoyed much wider impact, and remain
a key element of today's Internet applications, because they
are easy for programmers to use and export a clean abstraction for
dealing with recovery; however, when combined with relational
semantics, providing transactional guarantees requires substantial
engineering in order to get both good steady-state performance and
complete crash-safety and recovery.  Indeed, high-volume, high-performance
database systems cost hundreds of thousands  of dollars to
deploy and maintain.  Separating applications into stateless logic
plus transactions simplifies recovery; we exploit
this property by attempting application-generic recovery for the
logic, and intend to push it further by specializing the state stores
used for other kinds of Internet service state, including session
state and persistent non-relational state such as user profiles.

The authors of~\cite{press_performability} define a
performability~\cite{meyer:performability} metric for interactive
Internet services that captures the decrease in throughput as they
operate under fault conditions.  Their work is an important step in
applying performability to this domain, although as presented, it does
not capture the effect of faulty operation on a typical end-user---how
likely is it that a typical user request will fail or be delayed, and by
how much?  

An important part of managing partial failures is isolation.  Indeed, a
significant use of VMware's software is fault isolation, and there is a
new research focus on using lightweight virtual machines primarily for
isolation.  Recent efforts include Denali~\cite{whitaker:denali}, which
provides an Intel~x86 isolation kernel that omits support for some
instructions in exchange for orders-of-magnitude lighterweight
operation; JanosVM~\cite{tullmann:janos}, which allows a single logical
Java virtual machine to be split among multiple OS processes; and
Luna~\cite{hawblitzel:luna} and the Sun MVM~\cite{sun:mvm}, which
improve isolation among ``tasks'' running in a single Java VM.

BASE~\cite{rodrigues:base} and BFT~\cite{castro:bft} try to detect and
correct what would otherwise be silently-wrong answers, e.g. due to data
corruption or a malicious adversary.  Their work is complementary to
ours and composes with it, though we note that session state corruption
errors such as we encountered would be difficult for these approaches to
find as well.

\section{Future Work}

We intend to study
the feasibility of using microreboots for software
rejuvenation~\cite{huang:swrejuvenation}, the preemptive reboot of
applications to stave off failure caused by software aging (e.g., due
to resource leaks).  By microrejuvenating one component at a time in a
rolling fashion, we may be able to rejuvenate the entire applications
without any downtime or failed requests.
Combined with recursive restarts~\cite{candea:RR},
a technique that selectively restarts system components either
reactively or proactively, we hope microreboots can
preserve the benefits of reboots while eliminating some of their
drawbacks.

Other research has focused on developing microreboot-safe storage
systems for session state~\cite{ling:ssm} and non-relational persistent
state such as user profiles and catalog data~\cite{huang:dstore}.  We
intend to integrate these subsystems with our prototype to realize a
complete three-tier Internet application in which every subsystem is
microrebootable.  We hope to then use micoreboots  to enforce simple,
predictable 
fault models, by coercing any component failure into a fast-recovering
crash failure.  Using microreboots aggressively in this fashion has
the potential to help with containment of faults and to improve
predictability of component behavior in the face of failures.

Finally, although in this paper we tried to identify the limitations of
\urb{}ing on unmodified J2EE Internet applications, our longer-term goal
is the development of design rules, tools, and building blocks for
writing componentized applications for which recovery is dramatically
simplified because they are completely \urb-safe---that
is, crash-only applications~\cite{candea:crash-only}.

\section{Conclusions}

%---------------------------------------------------------------------------

We described some problems faced by Internet sites for which rebooting
in response to failure may be appealing but too expensive, and
proposed microreboots as a way to alleviate the problem.
Microreboot-based recovery works well for componentized applications
when the components have well-defined boundaries, when
externally-visible (to the component) state is either discardable or
persisted elsewhere, and when normal control flow mechanisms are
designed to handle the case of a component being temporarily
unavailable because it is recovering.  These constraints are largely
met by existing component architectures; we used a J2EE three-tier
application to show that microreboots can improve recovery time and
simplify recovery from transient faults, with no application-specific
\emph{a priori} knowledge.

Because microreboots are predictable and lightweight, it becomes
acceptable to make occasional mistakes regarding recovery, which in
turn enables the future use of sensitive fault detection techniques
for aggressive triggering of recovery.  This approach blurs the line
between normal operation and recovery and leads toward the design of
systems that are always ``recovering'' as one way of adapting to
changing conditions.  We hope that this will result in one concrete
path to ``self-*'' systems.

The current release of our software is available for download at
http://crash.stanford.edu/download; future versions will be posted
there as well.

%---------------------------------------------------------------------------

\bibliographystyle{../../ieee}
\bibliography{../rr}

\end{document}